\documentclass[smallextended]{svjour3}       
\smartqed  
\usepackage[pdftex]{graphicx}
\usepackage{amsmath,amsfonts,amssymb} 
 \journalname{AAPPS Bulletin}
\begin{document}

\title{Implication of Pulsar Timing Array Experiments on Cosmological Gravitational Wave Detection
}

\titlerunning{Implication of PTA experiments}        

\author{Jun'ichi Yokoyama
}


\institute{Research Center for the Early Universe (RESCEU) and Department of Physics,            Graduate School of Science, The University of Tokyo, Tokyo 113-0033, Japan
}

\date{RESCEU-9/21}

\maketitle

\begin{abstract}
 Gravitational waves provide a new probe of the Universe which can reveal a
number of cosmological and astrophysical phenomena that cannot be observed by
electromagnetic waves. Different frequencies of gravitational waves are
detected by different means. Among them, precision measurements of pulsar
timing provides a natural detector for gravitational waves with light-year
scale wavelengths. In this review, first a basic framework to detect a
stochastic gravitational wave background using pulsar timing array is
introduced, and then possible interpretations of the latest observational
result of 12.5 year NANOGrav data are described.
\end{abstract}

\section{INTRODUCTION}

Gravitational waves are ripples of spacetime first predicted by Einstein in his theory of general relativity in 1916. In Newtonian theory of gravitation and mechanics, both space and time are rigid in the sense that they are not affected by any material content existing in the Universe, and gravity is a nonlocal force. In contrast, Einstein advocated that matter and energy make the spacetime curved and the curvature itself is the essence of gravitation. 

In his theory a massive body curves the surrounding space and another particle moves in the curved space approaching the former as if it is directly attracted by the body. This is in close analogy of the electromagnetism of Faraday and Maxwell, where a moving charged particle produces electric and magnetic fields in the surrounding space, and another charged particle feels a force from these fields. In the case of gravity, the spacetime curvature is characterized by a metric tensor, ${g_{\mu \nu }}(x)$, with which the spacetime interval is expressed as $d{s^2} = {g_{\mu \nu }}(x)d{x^\mu }d{x^\nu }$ where repeated Greek indices are summed over spacetime coordinates 0 to 3. Without any gravitational effect, ${g_{\mu \nu }}(x)$ coincides with the Minkowski metric ${\eta _{\mu \nu }} = \mathrm{diag}( - 1,1,1,1)$, taking the speed of light unity.

Gravitational waves can be expressed by a small perturbation around the Minkowski metric in terms of the transverse-traceless gauge as
\begin{equation}
d{s^2} = - d{t^2} + \left( {{\delta _{ij}} + h_{ij}^{TT}(t,{\mathbf{x}})} \right)d{x^i}d{x^j} \label{1}
\end{equation}
which satisfies ${\partial _i}h_j^{TTi} = 0,{\ }h_i^{TTi} = 0$. Here repeated Latin indices are summed over the spatial coordinates 1 to 3. In the linearized weak-gravity limit of general relativity, the tensor perturbation $h_{ij}^{TT}(t,{\mathbf{x}})$ satisfies the same wave equation as the electromagnetic field, which is also equivalent to the Klein-Gordon equation for a massless scalar field, so that the gravitational wave propagates with the speed of light. 

One can easily see that propagating gravitational waves do not change the position of whatever body initially at rest whose motion obeys the geodesic equation: 
\begin{equation}
\frac{{{d^2}{x^\mu }}}{{d{\tau ^2}}} + \Gamma _{\nu \sigma }^\mu \frac{{d{x^\nu }}}{{d\tau }}\frac{{d{x^\sigma }}}{{d\tau }} = 0,{~~}\Gamma _{\nu \sigma }^\mu = \frac{1}{2}{g^{\mu \rho }}\left( {{\partial _\nu }{g_{\rho \sigma }} + {\partial _\sigma }{g_{\nu \rho }} - {\partial _\rho }{g_{\nu \sigma }}} \right)
\end{equation} 
where $\tau$ is the proper time. Since $\Gamma _{00}^i = 0$ for the metric (1), a star or a particle initially at rest does not feel any acceleration by the tensor perturbation and so remains at rest. Thus the only effect gravitational wave causes is the change of the light path or distance between two objects at rest. 

In the case of a ground-based detector such as the advanced LIGO (aLIGO)  \cite{1} in USA, advanced Virgo (aVirgo)  \cite{2} in Europe, and KAGRA  \cite{3} in Japan, two reflective laser lines are set with a rectangular configuration whose interference measures modulations of their relative path length to detect gravitational waves. Their baseline is 3-4 kilometers long and these detectors are sensitive to gravitational waves in deca- to hecto-Hertz band. 

While they were originally designed to detect gravitational wave event from binary neutron star coalescence, the first direct detection of gravitational waves, which was achieved by aLIGO on September 17, 2015, was from coalescence of two massive black holes around 30 solar mass \cite{4}. Up to then, even the existence of such a massive black hole had not been appreciated in the community. This manifests another example of common practice of the history of astronomy that every time the mankind obtained a new means of observation, a new unexpected event had been discovered. 

The first discovery of gravitational waves from binary neutron star coalescence was made on August 17, 2017  \cite{5} which led to joint observation of multiband photons ranging from gamma-ray to infrared \cite{6}. We note that this epoch making discovery was successfully made only by the GstLAL pipeline jointly developed by Kipp Cannon at RESCEU and Chad Hanna at Penn State University among the six pipelines of gravitational wave burst detection at work.

In order to detect lower frequency gravitational waves by a laser interferometer, we must extend the baseline, which may be achieved by launching a set of three satellite to form a laser interferometer of triangular configuration. This idea was first put into serious investigation by the Laser Interferometer Space Antenna (LISA) project which aims at launching three spacecrafts separated by 2.5 million km in a triangular formation orbiting around the Sun following the Earth in the same orbit \cite{7}. It will have the best sensitivity at milli-Hertz frequency range and can probe black holes with much larger masses than those discovered by aLIGO and aVirgo in the range $10^{2-7}$ solar mass. The path-finder satellite was launched in December 3, 2015 which was one day after the centennial of the publication of the Einstein equation  \cite{8}. The satellite exhibited much better performance than originally planned providing a hope to launch the full mission in 2034  \cite{9}. 

DECIGO in Japan, on the other hand, sets a goal of the direct detection of the stochastic gravitational wave background which was produced quantum mechanically during inflation in the early Universe, observing a decihertz range of gravitational waves  \cite{10}. In fact, DECIGO will ultimately be able to measure the reheating temperature after inflation which indicates when the Big Bang happened  \cite{11}. While it has been shown that the baseline should be as long as 1500km with a Fabry-Perot cavity  \cite{12}, which induces a resonance of laser by reflecting the light many times by mirrors at both ends of the cavity, current plan under discussion is to realize its degraded version called B-DECIGO whose baseline is only 150km first which is still in a phase of conceptual study.

In contrast, Chinese space-based projects are advancing much more rapidly with two independent projects of space-based gravitational wave detection, namely, Taiji  \cite{13} and TianQin  \cite{14}. Both projects have successfully launched the initial path finder satellites already.

The scope of this article is to focus on a set of gravitational wave detectors with by far the longer baseline prepared by nature, namely, timing observation of millisecond pulsars, or pulsar timing arrays.

Pulsars are rotating neutron stars with a strong magnetic field whose axis is not identical to the rotation axis, so that pulse-like periodic electromagnetic radiation is observed with the rotational period. So far about 2.5 thousand pulsars have been discovered since its first discovery in 1967 by Bell and Hewish  \cite{15}, and their rotation period spans from a millisecond to about ten seconds. Among them, the rotation period of millisecond pulsars with their period less than 30 msec is especially stable, most of which exist in a binary system and have a relatively weak magnetic field of order of $10^8$ G  \cite{16}. They are so stable that by observing possible modulation of the arrival times of their pulses, we can in principle detect stochastic gravitational wave background using observational timing data of many 
pulsars.
\begin{figure}
\includegraphics[width=10cm]{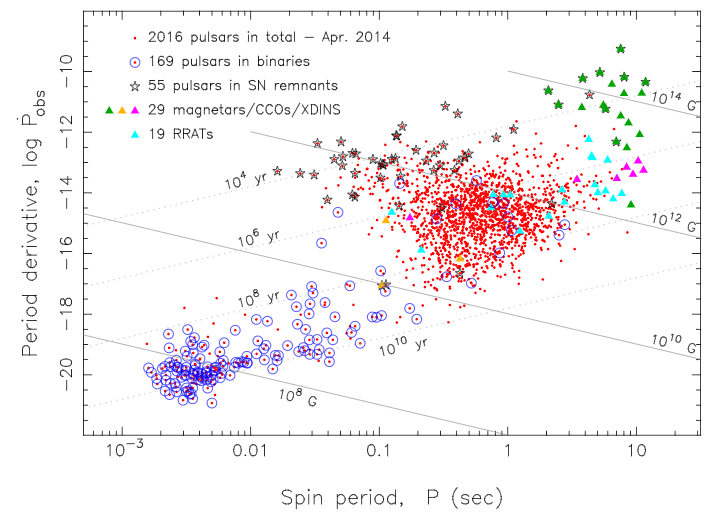}
\caption{ Distribution of rotation period and its time variation rate of known pulsars together with the contours of characteristic age and magnetic field. Taken from  \cite{16}.}
\label{fig:1} 
\end{figure}

Let us first summarize how the time of arrival (TOA) of the pulsar’s signal is modulated due to the propagating gravitational waves. For more detailed introduction, one may consult e.g.  \cite{17}.

\section{DETECTION OF GRAVITATIONAL WAVES USING PULSAR TIMING}

\subsection{Effects of gravitational waves on the timing of a single pulsar}
Suppose that an observer is at rest at the origin of the spacetime (1) and receive a signal emitted at ${t_{em}}$ from the a-th pulsar at the distance ${d_a}$ and directional cosine ${{\mathbf{n}}_a}$. The observer receives the signal at 
\begin{equation}
{t_{obs}} = {t_{em}} + {d_a} + \frac{{n_a^in_b^j}}{2}\int\limits_{{t_{em}}}^{{t_{em}} + {d_a}} {dt'} h_{ij}^{TT}[t',({t_{em}} + {d_a} - t'){{\mathbf{n}}_a}]. \label{2}
\end{equation} 
Repeating the same calculation one cycle $({T_a})$ later and subtracting (2), the modulation of pulsar period observed is given by
\begin{equation} 
\frac{{\Delta {T_a}}}{{{T_a}}} = \frac{{n_a^in_b^j}}{2}\int\limits_{{t_{em}}}^{{t_{em}} + {d_a}} {dt'} \frac{\partial }{{\partial t'}}h_{ij}^{TT}[t',{\mathbf{x}}], \label{3}
\end{equation} 
where the integral is calculated along the unperturbed photon path ${\mathbf{x}} = ({t_{em}} + {d_a} - t'){{\mathbf{n}}_a}$.

For a plane gravitational wave propagating along the direction ${\mathbf{n}}$,
$h_{ij}^{TT}(t,{\mathbf{x}}) \equiv {A_{ij}}({\mathbf{n}})\cos \left[ {\omega \left( {t - {\mathbf{n}} \cdot {\mathbf{x}}} \right)} \right]$,
we find 
\[
\frac{{\Delta {T_a}}}{{{T_a}}} = \frac{{n_a^in_a^j{A_{ij}}({\mathbf{n}})}}{{2(1 + {\mathbf{n}} \cdot {{\mathbf{n}}_a})}}\left\{ {\cos \left[ {\omega {t_{obs}}} \right] - \cos \left[ {\omega \left( {{t_{em}} - {\tau _a}{\mathbf{n}} \cdot {{\mathbf{n}}_a}} \right)} \right]} \right\},\]
with ${\tau _a} \equiv {t_{obs}} - {t_{em}}$. In general, it can be expressed as
\begin{equation}{z_a}(t) \equiv \frac{{\Delta {T_a}}}{{{T_a}}} = \frac{{n_a^in_b^j}}{{2(1 + {\mathbf{n}} \cdot {{\mathbf{n}}_a})}}\left\{ {h_{ij}^{TT}[t,{\mathbf{x}} = {\mathbf{0}}] - h_{ij}^{TT}[t - {\tau _a},{{\mathbf{x}}_a}]} \right\}. \label{4}
\end{equation}
The first term in the right-hand side is called the earth term and the second term the pulsar term. In the actual data analysis, the earth term is calculated at the barycenter of the solar system to remove seasonal and other modulations associated with the dynamics of the solar system.

Choosing the coordinate frame so that the gravitational wave propagates along the z direction, ${\mathbf{n}} = (0,0,1)$, $h_{ij}^{TT}$ reads
\begin{equation}
h_{ij}^{TT}(t - z) = \left( {\begin{array}{*{20}{c}}
{{h_ + }}&{{h_ \times }}&0 \\ 
{{h_ \times }}&{ - {h_ + }}&0 \\ 
0&0&0 
\end{array}} \right). \label{5}
\end{equation} 
If the a-th pulsar is located in the angular direction in terms of the polar coordinate, we find
\begin{align}
\frac{{\Delta {T_a}}}{{{T_a}}} \equiv {z_a}(t) = \frac{1}{2}&(1 - \cos {\theta _a})\left\{ {\cos 2} \right.{\phi _a}\left[ {{h_ + }(t) - {h_ + }(t - {\tau _a} - {\tau _a}\cos {\theta _a})} \right] \nonumber \\
&+ \sin 2{\phi _a}\left. {\left[ {{h_ \times }(t) - {h_ \times }(t - {\tau _a} - {\tau _a}\cos {\theta _a})} \right]} \right\}.
\end{align}
The dependence on $2{\phi _a}$ manifests the spin 2 nature of the graviton. 

We now apply it to the case of isotropic stochastic background of gravitational wave radiation where $h_{ij}^{TT}({\mathbf{x}},t)$is expressed in terms of the stochastic mode function ${\tilde h_A}(f,{\mathbf{n}})$ as
\begin{equation}
h_{ij}^{TT}(t,{\mathbf{x}}) = \sum\limits_{A = + , \times } {\int\limits_{ - \infty }^\infty {df\int {d{\Omega _{\mathbf{n}}}} } } {\tilde h_A}(f,{\mathbf{n}})e_{ij}^A({\mathbf{n}}){e^{ - 2\pi if(t - {\mathbf{n}} \cdot {\mathbf{x}})}} \label{6}
\end{equation} 
where $e_{ij}^A({\mathbf{n}})$ is the symmetric polarization tensor which satisfies ${n^i}e_{ij}^A({\mathbf{n}}) = 0$ and $e_{ij}^A({\mathbf{n}})e_{}^{A'ij}({\mathbf{n}}) = 2{\delta ^{AA'}}$.
We will return to their explicit form shortly. The stochastic background is characterized by the power spectrum ${S_h}(f)$ which satisfies
\begin{equation}
\left\langle {\tilde h_A^ * (f,{\mathbf{n}}){{\tilde h}_{A'}}(f',{\mathbf{n'}})} \right\rangle = \frac{1}{2}{S_h}(f)\delta (f - f')\frac{{\delta ({\mathbf{n}},{\mathbf{n'}})}}{{4\pi }}{\delta _{AA'}}. \label{7} \end{equation} 
The modulation of pulsar's period is then expressed as
\begin{equation}
{z_a}(t) = \sum\limits_{A = + , \times } {\int\limits_{ - \infty }^\infty {df\int {d{\Omega _{\mathbf{n}}}} } } {\tilde h_A}(f,{\mathbf{n}})F_a^A({\mathbf{n}})\left[ {1 - {e^{ - 2\pi if{\tau _a}(1 + {\mathbf{n}} \cdot {{\mathbf{n}}_a})}}} \right] \label{8}, \end{equation} 
with 
\[ F_a^A({\mathbf{n}}) = \frac{{n_a^in_a^je_{ij}^A({\mathbf{n}})}}{{2(1 + {\mathbf{n}} \cdot {{\mathbf{n}}_a})}}.\]
 The first term in the square bracket represents the earth term of (\ref{4}). The primary quantity of observational importance is the time integral of (\ref{4}) and (\ref{8}), ${r_a}(t) \equiv \int\limits_{{t_{ini}}}^t {{z_a}(t')dt'} $, which is called the timing residual.

\subsection{Effects of gravitational waves on the timing of a pair of pulsars}
In order to detect stochastic gravitational wave background in terms of pulsar timing array, it is important to study the correlation of the timing residual of two or more pulsars. The primary quantity is the equal-time correlation function of the timing residual of pulsars a and b, 
\begin{equation}\left\langle {{z_a}(t){z_b}(t)} \right\rangle = \frac{1}{2}\int\limits_{ - \infty }^\infty {df{S_h}(f)\int {\frac{{d{\Omega _{\mathbf{n}}}}}{{4\pi }}} } {K_{ab}}(f,{\mathbf{n}})\sum\limits_{A = + , \times } {F_a^A} ({\mathbf{n}})F_b^A({\mathbf{n}})\label{9}\end{equation} 
with ${K_{ab}}(f,{\mathbf{n}}) \equiv \left[ {1 - {e^{ - 2\pi if{\tau _a}(1 + {\mathbf{n}} \cdot {{\mathbf{n}}_a})}}} \right]\left[ {1 - {e^{ - 2\pi if{\tau _b}(1 + {\mathbf{n}} \cdot {{\mathbf{n}}_b})}}} \right]$which is well approximated by unity as the other terms are highly oscillatory. Let us express the directional vector of gravitational waves as ${\mathbf{n}} = (\sin \theta \cos \phi ,\sin \theta \sin \phi ,\cos \theta )$. 

Then introducing orthogonal vectors ${\mathbf{u}} = (\sin \phi , - \cos \phi ,0)$
and \\ ${\mathbf{v}} = (\cos \theta \cos \phi ,\cos \theta \sin \phi , - \sin \theta )$, we can express the polarization tensors as
\begin{align}
e_{ij}^ + ({\mathbf{n}}) = {{\mathbf{u}}_i} \otimes {{\mathbf{u}}_j} - {{\mathbf{v}}_i} \otimes {{\mathbf{v}}_j} \hfill \\
e_{ij}^ \times ({\mathbf{n}}) = {{\mathbf{u}}_i} \otimes {{\mathbf{v}}_j} + {{\mathbf{u}}_j} \otimes {{\mathbf{v}}_i} \hfill 
\end{align} 
with which we find
\begin{equation}
F_a^ + ({\mathbf{n}}) = \frac{{{{\left( {{{\mathbf{n}}_a} \cdot {\mathbf{u}}} \right)}^2} - {{\left( {{{\mathbf{n}}_a} \cdot {\mathbf{v}}} \right)}^2}}}{{2(1 + {\mathbf{n}} \cdot {{\mathbf{n}}_a})}},{~}F_a^ \times ({\mathbf{n}}) = \frac{{\left( {{{\mathbf{n}}_a} \cdot {\mathbf{u}}} \right)\left( {{{\mathbf{n}}_a} \cdot {\mathbf{v}}} \right)}}{{2(1 + {\mathbf{n}} \cdot {{\mathbf{n}}_a})}}\label{10}. \end{equation}

Thanks to the assumed isotropy of the stochastic background, (\ref{9}) depends only on the relative angle between pulsars a and b, ${\theta _{ab}}$. Then we may choose the z axis along the direction to the a-th pulsar, so that ${{\mathbf{n}}_a} = (0,0,1)$ and take ${{\mathbf{n}}_b} = (\sin {\theta _{ab}},0,\cos {\theta _{ab}})$. Inserting them to (\ref{9}) to calculate (\ref{8}) we find
\begin{equation}
\left\langle {{z_a}(t){z_b}(t)} \right\rangle = C\left( {{\theta _{ab}}} \right)\int\limits_0^\infty {df{S_h}(f)}, \label{11} \end{equation} 
where
\begin{equation}C\left( {{\theta _{ab}}} \right) \equiv \sum\limits_{A = + , \times } {\int {\frac{{d{\Omega _{\mathbf{n}}}}}{{4\pi }}} F_a^A} ({\mathbf{n}})F_b^A({\mathbf{n}}) = {x_{ab}}\ln {x_{ab}} - \frac{1}{6}{x_{ab}} + \frac{1}{3}\label{12}\end{equation} 
with ${x_{ab}} \equiv {\sin ^2}\tfrac{{{\theta _{ab}}}}{2}$. The most important feature of (11) is the fact that the form of the angular part is independent of the spectrum of gravitational radiation, ${S_h}(f)$,which was first found by Hellings and Downs  \cite{18} and the functional form of $C\left( {{\theta _{ab}}} \right)$ is often called the Hellings-Downs curve. 
\begin{figure}
\includegraphics[width=10cm]{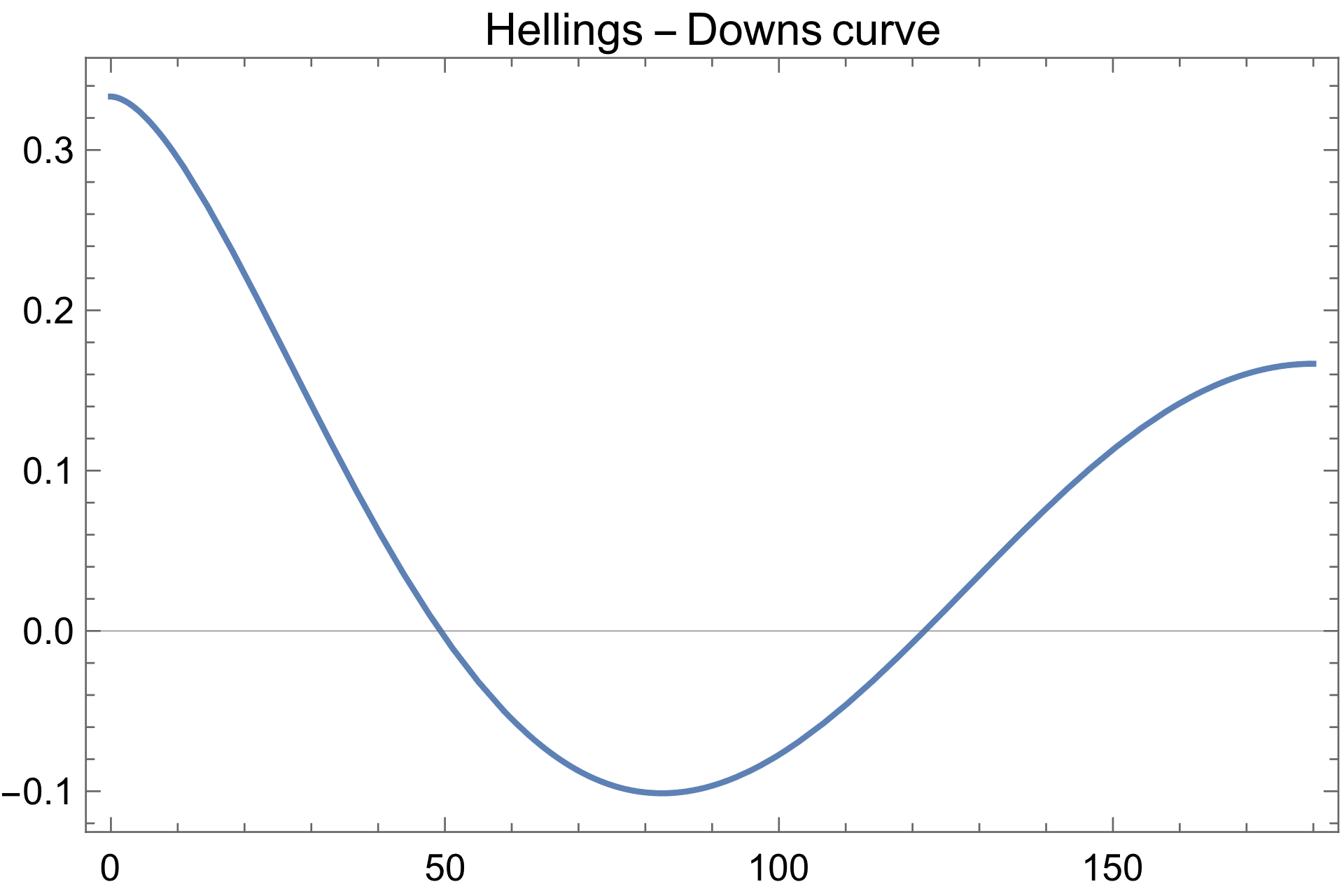}
\caption{ Hellings-Downs curve $C\left( {{\theta _{ab}}} \right)$ as a function of ${\theta _{ab}}$ [degree].}
\label{fig:2} 
\end{figure}

The correlation function of the timing residual reads 
\begin{equation}
\left\langle {{r_a}(t){r_b}(t)} \right\rangle = C\left( {{\theta _{ab}}} \right)\int\limits_0^\infty {df\frac{{{S_h}(f)}}{{{{\left( {2\pi f} \right)}^2}}}} \times 2\left[ {1 - \cos \left( {2\pi f(t - {t_{ini}})} \right)} \right]\label{13}. \end{equation}
We can extend the above to the case of unequal time but the result is always proportional to the Hellings-Downs function $C\left( {{\theta _{ab}}} \right)$ which reflects the quadrupole nature of gravitational radiation.

\subsection{Characteristic strain and density parameter of gravitational wave background}
The characteristic strain, ${h_c}(f)$, is defined by 
\begin{equation} h_c^2(f) \equiv 2f{S_h}(f)\label{14}. \end{equation} 
It is often modeled by a single power law:
\begin{equation}{h_c}(f) = {A_{GW}}{\left( {\frac{f}{{{f_{{\text{yr}}}}}}} \right)^\alpha }\label{15} \end{equation}
where it is convenient to take the characteristic frequency at ${f_{{\text{yr}}}} = 1{\text{y}}{{\text{r}}^{ - 1}}$when we analyze observation using pulsar timing.

To discuss cosmological implication of stochastic gravitational wave background, it is convenient to express the energy density of gravitational waves, which is naturally defined by
\begin{equation}{\rho _{gw}}(t) = \frac{1}{{32\pi G}}\left\langle {{{\dot h}_{ij}}{{\dot h}^{ij}}} \right\rangle ,\label{16} \end{equation} 
in terms of its contribution to the density parameter per logarithmic frequency interval, ${\Omega _{gw}}(f)$. Since (\ref{16}) is expressed as
\begin{equation}{\rho _{gw}}(t) = \frac{4}{{32\pi G}}\int\limits_0^\infty {df{{(2\pi f)}^2}} {S_h}(f) = \frac{\pi }{{2G}}\int {d\ln f{~}{f^3}} {S_h}(f)\label{17}, \end{equation}
${\Omega _{gw}}(f)$ reads
\begin{equation}{\Omega _{gw}}(f) \equiv \frac{1}{{{\rho _{cr}}}}\frac{{d{\rho _{gw}}(f)}}{{d\ln f}} = \frac{{4{\pi ^2}}}{{3H_0^2}}{f^3}{S_h}(f) = \frac{{2{\pi ^2}}}{{3H_0^2}}{f^2}h_c^2(f)\label{18}, \end{equation}
where ${\rho _{cr}} = {{3H_0^2} \mathord{\left/
{\vphantom {{3H_0^2} {(8}}} \right.
\kern-\nulldelimiterspace} {(8}}\pi G)$ is the critical density with ${H_0}$being the current Hubble parameter.

\subsection{Observables and data analysis }
The observed quantity is a time sequence of timing residuals of all the observed pulsars denoted as ${r_a}({t_i}),{~}a = 1,2,...{N_p};{~}i = 1,2,3,...$with ${N_p}$being the total number of pulsars regularly monitored. The observables may be decomposed into the deterministic part and noise part as
\begin{equation}{r_a}({t_i}) = r_a^{\det }({t_i},{{\mathbf{\xi }}_a}) + {n_a}({t_i})\label{19}\end{equation} 
where collectively denotes the parameters characterizing the nature of the a-th pulsar and other deterministic parameters modeling the observational setup  \cite{19}.

Assuming that noises are Gaussian distributed, we can perform statistical analysis once the two-point correlation function of noises is obtained. The auto-correlation function of noises of the same pulsar, ${N_{aa;ij}} = \left\langle {{n_a}({t_i}){n_a}({t_j})} \right\rangle $, consists of a frequency- independent white noise part such as instrumental errors and frequency dependent red noise term with excess power at lower frequencies such as spin noise, pulse profile changes, and imperfectly modeled dispersion measure variations. The stochastic gravitational wave background also contributes to the latter. Noises acting on different pulsars at different time, ${N_{ab;ij}} = \left\langle {{n_a}({t_i}){n_b}({t_j})} \right\rangle $, is subdominant compared with ${N_{aa;ij}}$but plays an important role to detect the stochastic background of gravitational waves, as their effect appear as its quadrupole term, while their monopole and dipole terms are mainly due to the errors in the reference clock on the earth and solar system modeling, respectively. 

In order to measure the amplitude of stochastic gravitational wave background, we first marginalize over the pulsar parameters ${{\mathbf{\xi }}_a}$assuming that gravitational waves are absent and then write down the likelihood function incorporating it. The log-likelihood function has a form
\begin{equation}\log P\left( {{\mathbf{\bar r}}|\left. {{A_{GW}}} \right)} \right. = - \frac{1}{2}{}^t{{\mathbf{\bar r}}_a}\bar N_{ab}^{ - 1}{{\mathbf{\bar r}}_b} - \frac{1}{2}\log \left[ {\det (2\pi \bar N)} \right]\label{20} \end{equation}
where ${{\mathbf{\bar r}}_a}$ and $\bar N$ collectively denotes the sequence of observed data with the deterministic part subtracted, ${r_a}({t_i}) - r_a^{\det }({t_i},{{\mathbf{\xi }}_a})$, and the covariance matrix of noises both after marginalization of deterministic parameters, respectively. 

We assume ${{\mathbf{\bar r}}_a}$and ${{\mathbf{\bar r}}_b}$with $a \ne b$has a vanishing correlation apart from the quadrupolar one due to the gravitational waves. We may express the marginalized noise covariance matrix as $\bar N = P + A_{GW}^2S$ where the matrix $S$ is estimated as ${S_{ab}} = \left\langle {{{{\mathbf{\bar r}}}_a}{}^t{{{\mathbf{\bar r}}}_b}} \right\rangle $with temporal coefficients being suppressed  \cite{20}. Since ${A_{GW}}$ is a tiny quantity if not zero, the inverse matrix is easily obtained as ${\bar N^{ - 1}} = {P^{ - 1}} - A_{GW}^2{P^{ - 1}}S{P^{ - 1}}$ to the lowest order in ${A_{GW}}$. We therefore find the log-likelihood ratio as
\begin{equation}\log \Lambda = \log \frac{{P\left( {{\mathbf{\bar r}}|\left. {{A_{GW}}} \right)} \right.}}{{P\left( {{\mathbf{\bar r}}|\left. {{A_{GW}} = 0} \right)} \right.}} = \frac{1}{2}A_{GW}^2{}^t{\mathbf{\bar r}}{P^{ - 1}}S{P^{ - 1}}{\mathbf{\bar r}}\label{21}. \end{equation}

We thus find the optimal statistic to calculate the amplitude of gravitational waves as
\begin{equation}{\hat A^2} \equiv \frac{{{}^t{\mathbf{\bar r}}{P^{ - 1}}S{P^{ - 1}}{\mathbf{\bar r}}}}{{{\text{tr}}\left[ {{P^{ - 1}}S{P^{ - 1}}S} \right]}}\label{22}, \end{equation}
which yields $\langle {{{\hat A}^2}} \rangle = A_{GW}^2$. If the signal is weak, the standard deviation is given by , so that the signal-to-noise ratio reads $S/N={\hat A}^2/ \sigma _0$     
\cite{21} \cite{22}.

In order to judge if nonvanishing stochastic gravitational wave background is present, one may adopt the Baysian analysis to obtain$P({M_i}|d)$, the probability of the model ${M_i}$ being correct for a given set of observational data ${\mathbf{d}}$. The Bayes theorem tells us
\begin{equation}\frac{{P({M_i}|{\mathbf{d}})}}{{P({M_j}|{\mathbf{d}})}} = \frac{{P({\mathbf{d}}|{M_i})}}{{P({\mathbf{d}}|{M_j})}}\frac{{P({M_i})}}{{P({M_j})}} \equiv {B_{ij}}\frac{{P({M_i})}}{{P({M_j})}} = {B_{ij}} \label{23} \end{equation}
where ${B_{ij}}$is the Bayes factor and the last equality applies for the flat prior case. If it is larger than unity the data prefers model $i$ to $j$, and its degree of preference has been classified by Jeffreys as follows. For ${\log _{10}}{B_{ij}} = 0 - 0.5$, the preference is not worth more than a bare mention, for ${\log _{10}}{B_{ij}} = 0.5 - 1$, it is substantial, for ${\log _{10}}{B_{ij}} = 1 - 2$, strong, and finally for ${\log _{10}}{B_{ij}} > 2$, it is decisive  \cite{23}.

As the observation continues, sensitivity to the gravitational wave background improves and we obtain more stringent upper bound on its amplitude with the frequency range extending toward lower frequencies. If there exist a nonvanishing background, we would first encounter stagnation of improvement in the upper limit, followed by emergence of spatially uncorrelated common-spectrum red noises to all the pulsars being monitored. Finally, we would find the quadrupolar signature in the spatial correlation described by the Hellings-Downs curve, which would serve as the final confirmation of the very existence of the gravitational wave background.

\section{LATEST 12.5yr OBSERVATION BY NANOGrav COLLABORATION}

\subsection{NANOGrav Observation}
There are three major pulsar timing array  \cite{24} (PTA) experiments worldwide, the European Pulsar Timing Array (EPTA)  \cite{25}, Parkes Pulsar Timing Array (PPTA)  \cite{26}, and the North American Nanohertz Observatory for Gravitational Waves (NANOGrav)  \cite{27}, which constitute the International Pulsar Timing Array (IPTA)  \cite{28}. Accumulation of observation for more than a decade now, the upper bound on the amplitude of the stochastic gravitational wave background has been improving to the frequency reaching . Recently, however, the 12.5 year NANOGrav data  \cite{29} has found a strong evidence of a common-spectrum stochastic process  \cite{30}.

The data consists of observation of 47 millisecond pulsars made from July 2004 to June 2017 by 305m Arecibo Observatory in Puerto Rico and 100m Green Bank Telescope in West Virginia  \cite{29}, and 45 pulsars observed for more than three years among them were used in the gravitational wave analysis. 

\subsection{Latest Result of NANOGrav Observation}
The 12.5year NANOGrav data set shows possible existence of common-spectrum process as depicted in Fig. 3 and fitted by a power law model in analogy with (15): 
\begin{equation}
{h_c}(f) = {A_{CP}}{\left( {\frac{f}{{{f_{{\text{yr}}}}}}} \right)^\alpha }~~{\rm with}~~{\gamma _{CP}} \equiv 3 - 2\alpha \label{24}. \end{equation}
Figures 3 and 4 indicate evidence for common red process at lower frequencies. Indeed the Bayes factor ($BF$) of the model with and without it reads $\log_{10}BF=4.5$ for the case the solar system ephemeris is fixed to DE436 provided by Jet Propulsion Laboratory and $\log_{10}BF=2.4$ for the case uncertainties in solar system ephemeris are marginalized using an algorithm called BayesEphem. The latter provides a more conservative result but in both cases, the existence of red noise is decisive. 

The next question is if the correlation among different pulsars satisfies the Hellings-Downs curve, or has monopolar or dipolar features. 
\begin{figure}[h]
\includegraphics[width=10cm]{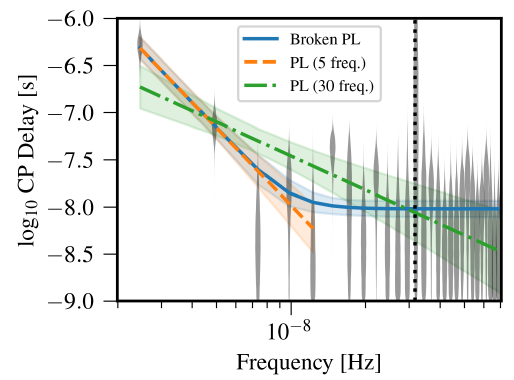}
\caption{Posterior spectrum of timing residual recovered from NANOGrav 12.5 year data with four models, namely, free spectrum (gray plots), broken power law (solid blue line), 5 lower frequency power-law fit (orange broken line), and 30 frequency power law fit (Green dot-dashed line). Dotted vertical line marks the frequency ${f_{{\text{yr}}}} = 1{\text{y}}{{\text{r}}^{ - 1}}$. Reproduced from ``The NANOGrav 12.5 yr Data Set: Search for an Isotropic Stochastic Gravitational-wave Background'' \cite{30} by permission of the AAS.}
\label{fig:3} 
\end{figure}

\begin{figure}[h]
\includegraphics[width=10cm]{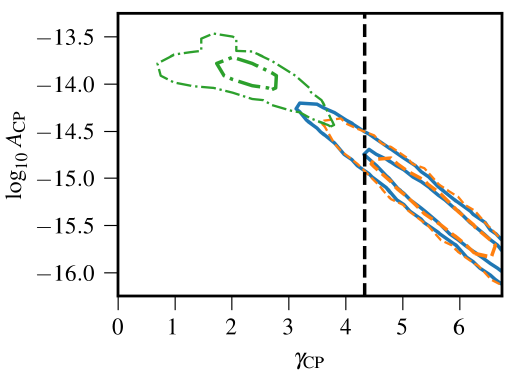}
\caption{ 1 and 2 $\sigma$ contours of posterior fit to the models described in Fig. 4. Green dot dashed curve is not a good model driven by higher frequency noises, while blue broken power-law contours refer to the fit at lower frequencies which is practically identical to the orange broken contours. The vertical broken line marks ${\gamma _{CP}} = {{13} \mathord{\left/
{\vphantom {{13} 3}} \right.
\kern-\nulldelimiterspace} 3}$corresponding to$\alpha = - {2 \mathord{\left/
{\vphantom {2 3}} \right.
\kern-\nulldelimiterspace} 3}$. Reproduced from \cite{30} by permission of the AAS.
}
\label{fig:4} 
\end{figure}

The monopolar correlation is strongly disfavored as $\log_{10}BF=-2.3$ with DE438 and $-1.3$ with BayesEphem. The dipolar correlation is also strongly disfavored as $\log_{10}BF=-2.4$ with DE438 and $-2.3$ with BayesEphem. On the other hand, the Bayes factor of the Hellings-Downs curve has been calculated as $\log_{10}BF=0.64$ with DE438 and 0.37 with BayesEphem, and mixtures with either common, monopole, or dipole process do not yield any higher BF. Hence it is premature to conclude that they have observed stochastic gravitational wave background yet. 
\begin{figure}[h]
\includegraphics[width=10cm]{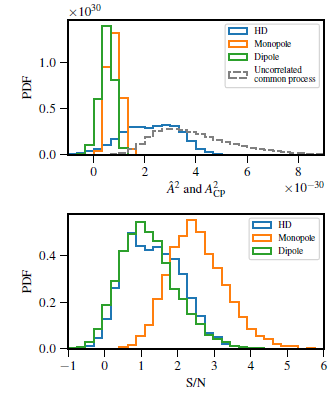}
\caption{Posterior distribution of the optimal statistic (upper panel) and S/N ratio (lower panel) for Hellings-Downs correlation (blue), monopolar (orange), and dipolar (green) angular correlations. Reproduced from \cite{30} by permission of the AAS.
}
\label{fig:5} 
\end{figure}

Figures 5 depict the posterior distributions of optimal statistic (16) for three types of correlations as well as the common-spectrum amplitude$A_{CP}^2$. Although the monopolar distribution shows somewhat higher S/N ratio, the central value is lower than the case of Hellings-Downs correlation, and so the result is consistent with the Bays factor analysis. 

Thus their result as a whole indicates a possible hint of the existence of stochastic gravitational wave background but by no means confirms it at this stage. We should still wait for the good news patiently. It is remarkable, however, a number of hasty theorists have already written many papers on many kinds of exotic interpretations of the result assuming that the result actually originates from gravitational wave background besides the more conventional possibility that it is from supermassive black hole binaries. In the next section we overview some of them in turn.

\section{POSSIBLE INTERPRETATIONS ASSUMING THE EXISTENCE OF GRAVITATIONAL WAVE BACKGROUND}

\subsection{Generation of gravitational waves}
Before discussing specific sources of gravitational wave background, let us introduce the generic quadrupole formula to provide the order-of-magnitude relation between the parameters of the source and the resultant amplitude of gravitational radiation. In this subsection we recover the dependence on the speed of light c. The linearized Einstein equation reads
\begin{equation}
 \Box h_{\mu \nu } = \left( { - \frac{1}{{c^2 }}\frac{{\partial ^2 }}{{\partial t^2 }} + \nabla ^2 } \right)h_{\mu \nu } = - \frac{{16\pi G}}{{c^4 }}T_{\mu \nu } , \label{25}
\end{equation} 
with being the energy-momentum tensor. The solution can easily be found with the retarded Green function as with the electromagnetism as
\begin{equation}
 h_{\mu \nu } (ct,{\bf{r}}) = \frac{{4G}}{{c^4 }}\int {\frac{{T_{\mu \nu } (ct - \left| {{\bf{r}} - {\bf{r'}}} \right|,{\bf{r'}})}}{{\left| {{\bf{r}} - {\bf{r'}}} \right|}}} d^3 r' , \label{26}
\end{equation}
Thanks to the conservation law, $
\partial _\mu T^{\mu \nu } = 0
$, one can express it in terms of the quadrupole moment as 
\begin{equation}
h_{ij} (ct,{\bf{r}}) = \frac{{2G}}{{c^4 r}}\ddot I(ct - r), \label{27}
\end{equation}
with
\begin{equation}
\ddot I(ct - r) = \int {\rho (ct,} {\bf{r'}})\left( {r'r'_j - \frac{1}{3}r'_k r'^k } \right)d^3 r'.
\label{28}
\end{equation} 
Thus in terms of the mass, $M$, size, $R$, time scale, $T$, of the source, we can express it as
\[
h(r) \sim \frac{G}{{c^4 r}}\frac{{MR^2 }}{{T^2 }} \le \frac{G}{{c^4 r}}Mc^2 \sim \frac{{GM}}{{c^2 r}} \sim \frac{{r_g }}{r},~~r_g \equiv \frac{{2GM}}{{c^2 }},
\]
where ${r_g}$ is the Schwarzschild radius of the source.

To discuss gravitational wave background of cosmological origin, the D'Alembert operator of (\ref{25}) must be replaced by the one in the Lemaitre-Robertson- Walker metric, and we must take relativistic effects in the energy momentum tensor into account to solve the Einstein equation instead of adopting slow-motion approximation used in (\ref{27}). 

\subsection{Binary supermassive black holes (SMBH)}
It is known that supermassive black holes exist ubiquitously in galactic nuclei  \cite{31} and in their evolution history it is expected that a number of coalescence took place of binary black holes. Their inspiral motion generates gravitational wave background which may be observed by pulsar timing  \cite{32}.  The spectrum of stochastic gravitational wave
background generated by their superposition was calcualted by Enoki et al \cite{32a} based on a semianalytic model of galaxy and quasar formation based on the hierarchical clustering scenario.

Suppose a pair of black holes with masses ${M_1}$ and ${M_2}$ forming a binary system in a circular orbit whose orbital radius is $R$. The frequency, ${f_r}$, of the gravitational waves emitted by this system is twice the orbital frequency, so from the Kepler’s law it satisfies $f_r^2 = {\pi ^2}GM{R^{ - 3}}$ where $M$ is the total mass. From the quadrupole formula (\ref{27}), we can estimate the amplitude of gravitational waves observed at a distance $r$ as
\begin{equation}h(r) \sim \frac{{G\mu }}{{{c^4}r}}{\left( {\frac{{GM}}{{f_r^2}}} \right)^{2/3}}f_r^2 \sim \frac{1}{r}{\left( {\frac{{G{M_c}}}{{{c^2}}}} \right)^{5/3}}{\left( {\frac{{{f_r}}}{c}} \right)^{2/3}}\label{29}\end{equation}
where $\mu $ is the reduced mass and ${M_c}$ is the chirp mass defined by ${M_c} \equiv {\mu ^{3/5}}{M^{2/5}}$. Then the power emitted by gravitational radiation scales as
\begin{equation}
P \sim \frac{{4\pi {r^2}}}{{16\pi G}}f_r^2{h^2}(r)c \sim \frac{1}{G}{\left( {\frac{{G{M_c}}}{{{c^2}}}} \right)^{10/3}}f_r^{10/3}\label{30}\end{equation} 
which also gives the energy loss rate of the binary system. Since the orbital energy is given by $E = - {{G{M_1}{M_2}} \mathord{\left/
{\vphantom {{G{M_1}{M_2}} {(2R)}}} \right.
\kern-\nulldelimiterspace} {(2R)}}$, we find $\dot E = P = - {{\dot RE} \mathord{\left/
{\vphantom {{\dot RE} {R = {{2{{\dot f}_r}E} \mathord{\left/
{\vphantom {{2{{\dot f}_r}E} {(3{f_r})}}} \right.
\kern-\nulldelimiterspace} {(3{f_r})}}}}} \right.
\kern-\nulldelimiterspace} {R = {{2{{\dot f}_r}E} \mathord{\left/
{\vphantom {{2{{\dot f}_r}E} {(3{f_r})}}} \right.
\kern-\nulldelimiterspace} {(3{f_r})}}}}$, so $E \propto f_r^{2/3}$and ${\dot f_r} \propto f_r^{11/3}$. We therefore find
\begin{equation}\frac{{dE}}{{d\ln {f_r}}} = f\frac{{\dot E}}{{{{\dot f}_r}}} = \frac{1}{3}{(G\pi )^{2/3}}M_c^{5/3}f_r^{2/3}\label{31}\end{equation} 

The observed frequency of gravitational waves emitted by a binary system at the redshift $z$ is $f = {(1 + z)^{ - 1}}{f_r}$, so the total energy density spectrum received by an observer today is 
\begin{equation}
\frac{{d{E_{gw}}(f)}}{{d\ln f}} = \int {\frac{{dN({M_c},z)}}{{d{M_c}dz}}\frac{{dE({f_r} = (1 + z)f)}}{{d\ln {f_r}}}\frac{{dz}}{{1 + z}}d{M_c}} \label{32}, \end{equation}
where $N({M_c},z)$ is the mass function of the black hole binary with the chirp mass${M_c}$. The left-hand side may be expressed as ${{\pi {c^2}{f^2}h_c^2(f)} \mathord{\left/
{\vphantom {{\pi {c^2}{f^2}h_c^2(f)} {(4G)}}} \right.
\kern-\nulldelimiterspace} {(4G)}}$ in terms of the characteristic strain$h_c^{}(f)$. Hence it has a dependence $h_c^{}(f) = A{f^{ - 2/3}}$, namely, $\alpha = - {2 \mathord{\left/
{\vphantom {2 3}} \right.
\kern-\nulldelimiterspace} 3}$  \cite{33} \cite{34}. 
\begin{figure}[h]
\includegraphics[width=10cm]{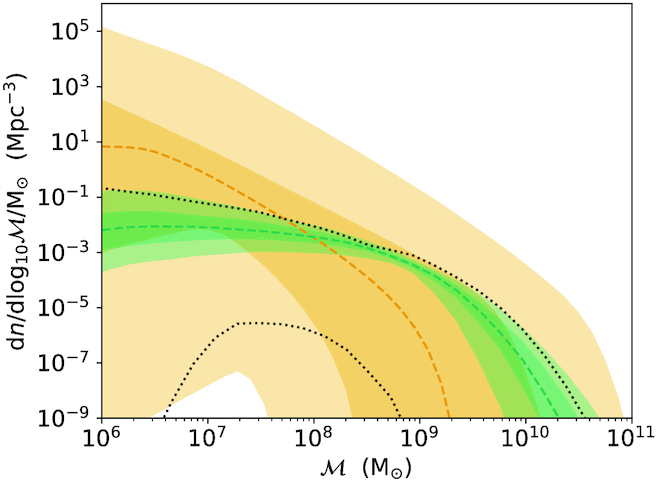}
\caption{The redshift-itegrated mass function of the chirp mass ${M_c} = \cal{M}$ of SMBHs inferred from 12.5 year NANOGrav observation. Dark and light orange (green) regions represent 50 and 90\% credible regions of M16 (C19) models. Taken from Figure 3 of 
``Massive black hole binary systems and the NANOGrav 12.5 yr results''  \cite{35}.
}
\label{fig:6} 
\end{figure}

This simple dependence is a nice feature of the binary SMBH scenario to identify the origin of the stochastic background on one hand, but this also means that we can extract information on their mass function only through the overall amplitude , which indicates it is difficult to translate the result to the differential mass spectrum in this simple treatment with circular orbits. 

In  \cite{35}, two models of SMBH evolution have been compared with the NANOGrav 12.5 year data. One is the model studied by Middleton et al (referred to M16), where the mass function of binary black hole system is assumed to be described by a Schechter function in terms of the chirp mass with the redshift evolution also having a Schechter type dependence on with different sets of parameters  \cite{36}. Their model has five parameters, namely, power index and exponential scale height for both chirp mass and the redshift to determine the shape of Schechter function as well as the overall amplitude. 

The other model studied by Chen et al (referred to C19) is more sophisticated as it incorporates eccentricity of the orbit and interaction with environment. It depends on 18 model parameters and has more complicated frequency dependence than a simple power law  \cite{37}. 

Figure 6  \cite{35} represents the redshift-integrated mass function of the SMBH chirp mass fitted to these two models, showing that the mass function in the simpler model with a single power-law frequency dependence is less constrained than the more sophisticated one.

\subsection{Cosmic strings}
The power index of the gravitational wave background inferred by 12.5 year NANOGrav is also consistent with smaller values than $\alpha = - {2 \mathord{\left/
{\vphantom {2 3}} \right.
\kern-\nulldelimiterspace} 3}$ which may be preferred by cosmological gravitational wave background.

Among them, cosmic strings are line-like topological defects created by a cosmological phase transition when the vacuum state after the symmetry breaking allows a nontrivial mapping to a unit circle such as the case U(1) symmetry is broken  \cite{38}. The simplest field theoretic model allowing a string solution is the Abelian Higgs model described by the Lagrangian with a complex scalar field $\Phi$,
\begin{align}
{\cal L} = {D_\mu }\Phi {({D^\mu }\Phi )^\dag } - \frac{1}{4}{F_{\mu \nu }}{F^{\mu \nu }} + {m^2}\Phi {\Phi ^\dag } - \lambda {(\Phi {\Phi ^\dag })^2} - \frac{\lambda }{4}{v^4} \hfill \\
{F_{\mu \nu }} = {\partial _\mu }{A_\nu } - {\partial _\nu }{A_\mu },{~}{D_\mu }\Phi = {\partial _\mu }\Phi - ig{A_\mu }\Phi ,{~}{m^2} = \lambda {v^2}. \hfill 
\end{align} 
The cosmic string is characterized by the dimensionless tension $G\mu $ where $\mu $is related to the symmetry breaking scale of the theory by $\mu = b\pi {v^2}$with $b$ being a function of ${{{g^2}} \mathord{\left/
{\vphantom {{{g^2}} \lambda }} \right.
\kern-\nulldelimiterspace} \lambda }$ \cite{39}. It takes a value between 0.5 and 3 as ${{{g^2}} \mathord{\left/
{\vphantom {{{g^2}} \lambda }} \right.
\kern-\nulldelimiterspace} \lambda }$ is varied from 100 to 0.01  \cite{40}.

Production of cosmic strings is usually attributed to a thermal phase transition which takes place due to the high-temperature correction to the potential  \cite{41}. The critical temperature of the above Abelian Higgs model is 
\begin{equation}{T_c} = 7 \times {10^{12}}{b^{ - {1 \mathord{\left/
{\vphantom {1 2}} \right.
\kern-\nulldelimiterspace} 2}}}{\left( {\frac{{G\mu }}{{{{10}^{ - 10}}}}} \right)^{{1 \mathord{\left/
{\vphantom {1 2}} \right.
\kern-\nulldelimiterspace} 2}}}{\text{GeV}}\label{33}\end{equation} 
which is fairly high compared with typical reheating temperature after cosmic inflation  \cite{42}. 

For models with lower reheating temperature, one can still produce strings after inflation if is nonminimally coupled to scalar curvature ${\cal R}$ with the term $\xi {\cal R}\Phi {\Phi ^\dag }$ \cite{43}. Then symmetry is restored during inflation if $12\xi {H^2} > \lambda {v^2}$ \cite{44}, or the tensor-to-scalar ratio satisfies
\begin{equation}
r > 4 \times {10^{ - 3}}\left( {\frac{\lambda }{{0.1}}} \right)\left( {\frac{\xi }{{{1 \mathord{\left/
{\vphantom {1 6}} \right.
\kern-\nulldelimiterspace} 6}}}} \right){b^{ - 1}}\left( {\frac{{G\mu }}{{{{10}^{ - 10}}}}} \right)\label{34}.\end{equation} 

It is known that 80\% of the energy of strings are in infinitely long ones and they evolve in the expanding universe intersecting with each other to form loops  \cite{45}. As a result they evolve according to the scaling solution in which there always exists fixed number of infinitely long strings in the horizon volume at each time and loops are formed with the constant comoving rate with the initial length proportional to the formation time as $\alpha t$. Numerical simulations show that has an extended spectrum up to $\alpha \approx 0.1$  \cite{46}. We take $\alpha = 0.1$ hereafter as corresponding loops make most important contributions to the gravitational wave background. Loops oscillate and decay by emitting gravitational waves with an energy-loss rate independent of its size, $\dot E = \Gamma G{\mu ^2}$, where $\Gamma $ is a constant around 50  \cite{47}. As a result, a loop formed at ${t_i}$ has a length
$\ell (t) = \alpha {t_i} - \Gamma G\mu (t - {t_i})$at $t$ and emits gravitational wave with a frequency 
$f(t) = {{2k} \mathord{\left/{\vphantom {{2k} {\ell (t)}}} \right.\kern-\nulldelimiterspace} {\ell (t)}}$, where $k$ is a positive integer. It is known that k-th harmonics has an emission rate proportional to 
${k^{{{ - 4} \mathord{\left/
{\vphantom {{ - 4} 3}} \right.
\kern-\nulldelimiterspace} 3}}}$ \cite{48}. 

Noting that the string loop contributing to the current gravitational wave frequency $f$ through k-mode emitted at time$t'$was created at
\begin{equation}{t_k}(t',f) = \frac{1}{{\alpha + \Gamma G\mu }}\left[ {\frac{{2k}}{f}\frac{{a(t')}}{{a({t_0})}} + \Gamma G\mu t'} \right]
\label{35}\end{equation} 
and summing up all the contribution throughout the cosmic evolution, we find the current spectrum of stochastic gravitational wave background
\begin{equation}{\Omega _{gw}}(f) \cong \sum\limits_k {\frac{{4\pi {k^{{1 \mathord{\left/
{\vphantom {1 3}} \right.
\kern-\nulldelimiterspace} 3}}}\Gamma {{(G\mu )}^2}}}{{27H_0^2f}}} \int\limits_{{t_f}}^{{t_0}} {dt'} \frac{{{C_{eff}}({t_k})}}{{t_k^4}}{\left( {\frac{{a(t')}}{{a({t_0})}}} \right)^5}{\left( {\frac{{a({t_k})}}{{a(t')}}} \right)^3}\Theta \left( {{t_k} - {t_f}} \right)\label{36}\end{equation} 
where ${t_f}$is the onset of the scaling regime. ${C_{eff}}$is a factor representing the dependence of the scaling solution on the cosmic expansion law  \cite{49}. We find ${C_{eff}} = 5.4(0.39)$ in radiation (matter) dominated regime.
\begin{figure}
\includegraphics[width=10cm]{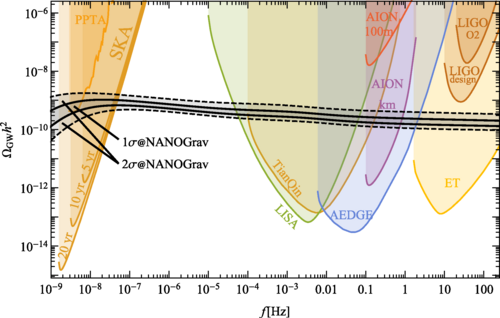}
\caption{Spectrum of stochastic gravitational wave back ground generated by cosmic string networks in the scaling regime fitted to the 12.5 year NANOGrav data assuming that their signal entirely comes from strings. Taken from  \cite{50}. 
}
\label{fig:7} 
\end{figure}

Figure 7 shows the energy spectrum of stochastic background fitted to the 12.5 year NANOGrav data assuming that they are entirely from cosmic string signals, where it is found that 68
If we allow variation of $\alpha $, even wider range of tension is consistent  \cite{51}. 

The stochastic background extended to higher frequencies as seen in Fig. 7 will cause big noise sources for the future space-based laser interferometers such as LISA, TianQin, Taiji, and DECIGO as well as to the Einstein Telescope  \cite{52}. Before reaching the era of the next and next-to-next detectors, however, once the world-wide network of currently available ground-based detectors, namely, aLIGO, aVirgo, and KAGRA (LVK) are in full operation with their respective design sensitivity, we can probe stochastic background to the level well below ${\Omega _{gw}}(f) < {10^{ - 9}}$ \cite{53} and touch the “NANOGrav curve” in Fig. 7. Thus the cosmic string scenario may be verified even with the current technology. 

Even if LVK did not observe the stochastic signal, however, it would not necessarily mean that cosmic string scenario would not work, because Fig. 7 is depicted under the assumption that waves relevant to LVK observation were created under radiation domination. If the reheating temperature after inflation is sufficiently low, they would have been created during the inflaton’s oscillation regime when the cosmic expansion law was the same as in the matter domination, and the energy spectrum would acquire suppression factor proportional to ${f^{ - 2}}$  \cite{54}. I estimate the upper bound on the reheating temperature is around 400GeV in the case NANOGrav observation is entirely due to cosmic strings but LVK in full operation does not find any stochastic background. 

\subsection{Tensor perturbation from inflation}
Cosmic inflation in the early universe is an indispensable ingredient of modern cosmology to realize big and old universe filled with hierarchical structures  \cite{42}. During inflation, each Fourier mode of tensor perturbation defined in (6), ${\tilde h_A}(f,{\mathbf{n}})$, satisfies the Klein-Gordon equation for a massless field, and can be quantized in the same way as a massless scalar field minimally coupled to gravity. As a result, it acquires nearly scale-invariant dispersion given by 
\begin{equation}
\frac{{4\pi {k^3}}}{{{{(2\pi )}^3}}}\left\langle {{{\left| {{{\tilde h}_A}(f,{\mathbf{n}})} \right|}^2}} \right\rangle \equiv {{\cal P}_T}(k) = 64\pi G{\left( {\frac{{H({t_k})}}{{2\pi }}} \right)^2}\label{37}, \end{equation}
where the pre-factor on the left-hand side is the phase space density with $k = 2\pi f$ and $64\pi G$originates from the normalization factor of the Einstein-Hilbert action through canonical quantization procedure  \cite{55}. The spectral index of the tensor perturbation is defined by
\begin{equation}
{n_t} \equiv \frac{{d{{\cal P}_T}(k)}}{{d\ln k}} = - 2{\varepsilon _H},{\text{ }}{\varepsilon _H} \equiv - \frac{{\dot H}}{{{H^2}}}\label{38}.
\end{equation} 
Here $H({t_k})$ is the Hubble parameter during inflation when the comoving wavenumber $k = 2\pi f$ left the Hubble horizon during inflation.

In the exact de Sitter background we find a scale-invariant spectrum with ${n_t} = 0$. In vast majority of inflation models ${n_t}$ takes a small negative value as ${n_t} > 0$ means violation of null energy condition. Since CMB imposes a stringent constraint on the amplitude of gravitational waves or tensor perturbations on large scales we need a positive tensor spectral index if we wish to account for the NANOGrav observation in terms of inflationary tensor perturbation. At the same time we must suppress higher frequency stochastic background as aLIGO has already obtained a constraint more stringent than that imposed by successful big bang nucleosynthesis (BBN), namely, ${\Omega _{gw}} < 6 \times {10^{ - 8}}$ for a flat spectrum  \cite{57}. This can be achieved by prolonged reheating phase after inflation with a low reheating temperature as mentioned in the last subsection. 

Parametric analysis blind to specific models has been done in  \cite{58} and we can read off range of parameters required to explain the NANOGrav data by focusing on the border line of pulsar constraints. As a result, the possible positive detection can be explained by an inflation model with a large tensor spectral index ${n_t} \approx 0.9$ and low reheating temperature ${T_R} \leq \cal{O}({10^2})$ GeV.

Such a blue tensor spectrum may be realized in G-inflation  \cite{59} \cite{60} \cite{61} or Gauss-Bonnet inflation  \cite{62}. 
Both of them can be analyzed using the generalized G-inflation framework  \cite{63}. A relatively simple model to explain 12.5year NANOGrav data by tensor perturbation of quantum origin has been proposed by Tahara and Kobayashi making use of only G2 and G3 generalized galileon functions  \cite{64}.

\subsection{Tensor perturbation generated by second-order scalar perturbations from inflation}
During inflationary expansion in the early universe, curvature perturbations are generated in a similar manner to tensor perturbation  \cite{65}. Although the simplest canonical slow-roll models predict nearly scale-invariant spectrum, nontrivial dependence of curvature perturbation on the scalar field model makes it possible to realize much larger amplitudes on smaller scale while preserving the consistency with CMB observations which determine the amplitude of the curvature perturbations of order of on large scale  \cite{56}. In particular, if slow-roll parameters satisfy a certain condition, the would-be decaying mode of the curvature perturbation can grow even after horizon exit, and can result in an enhanced spectrum on some specific scales  \cite{66}. If this happens, although tensor and scalar perturbations are decoupled at the level of linear perturbation, second-order effect may generate an appreciable amplitude of tensor perturbations  \cite{67} \cite{68}. This feature, pioneered in  \cite{69} \cite{70}, has been extensively discussed recently in relation to the formation of primordial black holes (PBHs)  \cite{71} \cite{72} which also requires high amplitude density perturbation of order of 0.1.

PBHs are produced when a region with a large density perturbation of order of unity falls in the horizon in the early universe dominated by radiation  \cite{73} \cite{74}. Its mass is of order of the horizon mass
\begin{equation}{M_{PBH}} \sim \frac{{{c^3}t}}{G} \sim {M_ \odot }\left( {\frac{t}{{{{10}^{ - 5}}\sec }}} \right)\label{39} \end{equation}
at formation. Spectrum of density perturbation with a peak realizing formation of PBHs with mass creates stochastic gravitational wave background at the typical frequency
\begin{equation}
f_{GW} \cong 9\left( {\frac{{M_{PBH} }}{{M_ \odot }}} \right)^{ - {1 \mathord{\left/
{\vphantom {1 2}} \right.
\kern-\nulldelimiterspace} 2}} {\rm{nHz}}
\label{40}. \end{equation}
Hence gravitational wave background of nano-hertz range is an important probe of PBHs around the solar mass. In particular, as the observation time gets longer, we will be able to constrain lower frequency background, so that we can eventually probe the mass range of the binary black holes observed by advanced LIGO and advanced Virgo.

The possible positive detection of the background by NANOGrav collaboration has somewhat affected this strategy as a number of authors have proposed to attribute it to the positive signature of the existence of PBHs. Among them, Kohri and Terada  \cite{75} argue that NANOGrav may suggest the presence of PBHs around the solar mass, while Luca, Franciolini, and Riotto  \cite{76} suggest a spectrum of curvature perturbation predicting PBHs with a wide range of masses which as a whole constitute all the dark matter. Vaskonen and Veermäe  \cite{77} take into account the effect of critical phenomenon on the mass function of PBHs and obtained a comprehensive contour constraining the average mass and the abundance of PBHs in terms of the amplitude and slope of the gravitational wave background. 

All these three works use a formula first obtained by Carr  \cite{79}, which is often called Press-Schechter type formula even though it is not based on their theory \cite{78},  to relate the abundance of PBHs and the amplitude of density perturbations.  It is desirable to reanalyze the problem under the improved mass function recently proposed  \cite{80} \cite{81} including the dependence on the choice of the window function  \cite{82} \cite{83}.

\subsection{First order phase transition}
Quantum field theory with spontaneous symmetry breaking may induce a cosmological phase transition in the early universe due to the high temperature or high curvature effects. If a first-order phase transition takes place, it provides several sources of gravitational waves, namely, collisions of true vacuum bubbles, sound waves, and turbulences  \cite{84}. In order to produce gravitational waves at nano-hertz frequencies, the phase transition must have taken at a temperature around 10-100MeV. In the standard model of particle physics, only QCD phase transition is relevant in this energy scale, but it is already known to be a cross over without appreciable production of gravitational radiation. Hence new theoretical input such as a dark sector inducing a dark phase transition is required  \cite{85} \cite{86}.

\subsection{Turbulence in magnetohydrodynamics} 
It is known that magnetohydrodynamics (MHD) turbulence in the early universe may induce stochastic gravitational wave background. For example, turbulence in the plasma and magnetic fields produced by a first-order phase transition at the electroweak scale may generate gravitational waves detectable by LISA  \cite{87}. In order to create gravitational waves relevant to NANOGrav observation in the same manner, we need a first-order QCD phase transition. Although such a possibility is interesting as we may relate the observation of gravitational waves with primordial magnetic fields, some new physics is required to realize such a phase transition  \cite{88}. 

\section{CONCLUSION}

In this article, I have briefly introduced a basic framework to observe gravitational waves by pulsar timing experiments and possible interpretations of the latest 12.5year observation of NANOGrav collaboration which reports a positive common spectrum signature that may be a hint of finite stochastic gravitational wave background in the nano-hertz frequency range. At present it is not guaranteed that the observed hint is due to the isotropic stochastic background of gravitational waves as the characteristic Hellings-Downs curve of angular correlation  \cite{18} has not been confirmed. Furthermore, most of the preferred amplitude of NANOGrav observation is in tension with the upper bound which has been obtained by PPTA, although they turn out to be consistent with each other allowing the uncertainties in ephemeris modeling.  Hence we should continue careful study both on observational and theoretical analyses  \cite{89}. 

\begin{acknowledgement}
The author wishes to thank Joe Simon, Vicky Kaspi, Alberto Sesana, and  Hannah Middleton
and their collaborators for allowing  to use their figures.  
This work is  supported by JSPS KAKENHI Grant JP20H05639 and Innovative Area JP20H05248.
\end{acknowledgement}

%
%


\begin{thebibliography}{}
%
 \bibitem{1} J. Aasi et al., Class. Quant. Grav. 32(2015)074001
 \bibitem{2} F. Acernese et al., Class. Quant. Grav. 32(2015) 024001.
 \bibitem{3} T. Akutsu et al., Nature Astron. 3(2019)35.
 \bibitem{4} B.P. Abbott et al., Phys. Rev. Lett. 116(2016)061102
 \bibitem{5} B.P. Abbott et al., LIGO  and  Virgo collaborations, Phys. Rev. Lett. 119(2017)161101
 \bibitem{6} B.P. Abbott et al., Astrophys. J. 848(2017)L13
 \bibitem{7} P. Amaro-Seoane et al. 1702.00786.
 \bibitem{8} P. McNamara, S. Vitale,  and  K. Danzmann, Class. Quant. Grav. 25(2008)114034
 \bibitem{9} M. Armano et al., Phys. Rev. Lett. 116(2016)231101
 \bibitem{10} S. Kawamura et al. Class. Quant. Grav. 28(2011) 094011.
 \bibitem{11} K. Nakayama, S. Saito, Y. Suwa, and J. Yokoyama, JCAP 06(2008)020.
 \bibitem{12} S. Kuroyanagi, K. Nakayama, and J. Yokoyama, PTEP 2015(2015)013E02.
 \bibitem{13} W.R. Hu  and  Y.L. Wu, Natl. Sci. Rev. 4(2007)685.
 \bibitem{14} J. Luo et al. Class. Quant. Grav. 33(2016)035010.
 \bibitem{15} A. Hewish, S.J. Bell, et al. Nature 217(1968)709.
 \bibitem{16} T.M. Tauris et al. PoS(AASKA14)039  (2015), arXiv 1501.00005.
 \bibitem{17} M. Maggiore, “Gravitaional Waves” vols 1 and 2, Oxford (2008, 2018).
 \bibitem{18} R.W. Hellings and G.S. Downs, Astrophys. J. 265(1983)L39.
 \bibitem{19} R. van Haasteren, Y. Levin, P. McDonald,  and  T. Lu, MNRAS 395(2009)1005.
 \bibitem{20} M. Anholm, S. Ballmer, J.D.E. Creighton, L.A. Price,  and  X. Siemens, Phys. Rev. D79(2009)084030.
 \bibitem{21} X. Siemens, J. Ellis, F. Janet,  and  J.D. Romano, Class. Quant. Grav. 30(2013)224015.
 \bibitem{22} S.J. Chamberlin et al., Phys. Rev. D91(2015)044048.
 \bibitem{23} H. Jeffreys, “Theory of Probability” Oxford (1961). 
 \bibitem{24} S. Detweiler, Astrophys. J. 234(1979)1100.
 \bibitem{25} G. Desvignes et al., MNRAS 458(2016)3341.
 \bibitem{26} D.J. Reardon et al., MNRAS 455(2016)1751.
 \bibitem{27} Z. Arzoumanian et al., Astrophys. J. Suppl. 235 (2018) 37.
 \bibitem{28} B.B.P. Perera et al., MNRAS 490(2019)4666.
 \bibitem{29} Md.F. Alam et al., Astrophysical J. Suppl 252(2021)4, arXiv 2005.06490.
 \bibitem{30} Z. Arzoumanian et al., Astrophys. J. Lett 905(2020)L34, arXiv 2009.04496.
 \bibitem{31} J. Kormendy  and  L.C. Ho, Ann. Rev. Astron. Astrophys. 51(2013)511.
 \bibitem{32} M. Rajagopal  and  R.W. Romani, Astrophys. J. 446 (1995)543.
 \bibitem{32a} M. Enoki, K.T. Inoue, M. Nagashima and N. Sugiyama, Astrophys. J. 615(2004)19.
 \bibitem{33} E.S. Phinney, (2001) astro-ph/0108028.
 \bibitem{34} A. Sesana, MNRAS 433(2013)L1.
 \bibitem{35} H. Middleton et al. MNRAS 502(2021)L99. 
 \bibitem{36} H. Middleton et al. MNRAS 455(2016)L72
 \bibitem{37} S. Chen, A. Sesana,  and  C.J. Conselice, MNRAS 488(2019)401.
 \bibitem{38} A. Vilenkin  and  E.P.S. Shellard, “Cosmic strings and other topological defects,” Cambridge (1994).
 \bibitem{39} H.B. Nielsen and P. Olesen, Nucl. Phys. B61 (1973)45.
 \bibitem{40} E.W. Kolb and M.S. Turner, “The Early Universe” CRC press (1990).
 \bibitem{41} T. Kibble, J. Phys. A9(1976)1387.
 \bibitem{42} K. Sato and J. Yokoyama, Int. J. Mod Phys. D24 (2015)1530025.
 \bibitem{43} J. Yokoyama, Phys. Lett. B 212(1988)273.
 \bibitem{44} J. Yokoyama, Phys. Rev. Lett. 63(1989)712.
 \bibitem{45} T. Vachaspati  and  A. Vilenkin, Phys. Rev. D30 (1984) 2036.
 \bibitem{46} J.J. Blanco-Pillado, K.D. Olum,  and B. Shlaer, Phys. Rev. D89(2014)023512.
 \bibitem{47} T. Vachaspati  and  A. Vilenkin, Phys. Rev. D31 (1985) 3052.
 \bibitem{48} J.J. Blanco-Pillado  and  K.D. Olum, Phys. Rev. D96(2017)104046.
 \bibitem{49} Y. Cui, M. Lewicki, D.E. Morrissey,  and  J.D. Wells, JHEP 01(2019)081.
 \bibitem{50} J. Ellis  and  M. Lewicki, Phys. Rev. Lett. 126(2021)041304.
 \bibitem{51} S. Blasi, V. Brdar,  and  K. Schmitz, 2009.06607.
 \bibitem{52} M. Maggiore et al., JCAP03(2020)050, arXiv 1912.02622.
 \bibitem{53} K. Schmitz, JHEP 01 (2021) 097.
 \bibitem{54} N. Seto  and  J. Yokoyama, J. Phys. Soc. Japan, 72 (2003) 3082. 
 \bibitem{55} A.A. Starobinsky, JETP Letters 30(1979)682.
 \bibitem{56} Planck collaboration, Y. Akrami et al. Astron.  and  Astrophys. 641(2020)A10.
 \bibitem{57} B.P. Abbott et al. Phys. Rev. D100(2019)061101.
 \bibitem{58} S. Kuroyanagi, T. Takahashi,  and  S. Yokoyama, JCAP 02(2015)003.
 \bibitem{59} T. Kobayashi, M. Yamaguchi,  and  J. Yokoyama, Phys. Rev. Lett. 105(2010)231302.
 \bibitem{60} Y.F. Cai et al. Nucl. Phys. B 900(2015)517.
 \bibitem{61} Y. Mishima  and  T. Kobayashi, Phys. Rev. D 101(2020)043536.
 \bibitem{62} S. Koh, B-H. Lee, and G. Tumurtushaa, Phs. Rev. D98 (2018)103511.
 \bibitem{63} T. Kobayashi, M. Yamaguchi,  and  J. Yokoyama, Prog. Theor. Phys. 126(2011)511.
 \bibitem{64} H.W.H. Tahara  and  T. Kobayashi, Phys. Rev. D 102 (2020)123533.
 \bibitem{65} V. Mukhanov  and  G. Chibisov, JETP Letters 33 (1981) 532.
 \bibitem{66} R. Saito, J. Yokoyama,  and  R. Nagata, JCAP 06 (2008) 024.
 \bibitem{67} K.N. Ananda, C. Clarkson, and D. Wands, Phys. Rev. D75(2007)123518.
 \bibitem{68} D. Baumann, P.J. Steinhardt, K. Takahashi, and K. Ichiki, Phys. Rev. D7(2007)084019
 \bibitem{69} R. Saito  and  J. Yokoyama, Phys. Rev. Lett. 102(2009)161101, 107(2011)069901(E).
 \bibitem{70} R. Saito  and  J. Yokoyama, Prog. Theor. Phys. 123(2010)867, 126(2011)351(E).
 \bibitem{71} B.J. Carr, K. Kohri, Y. Sendouda,  and  J. Yokoyama, 2002.12778.
 \bibitem{72} B.J. Carr, K. Kohri, Y. Sendouda,  and  J. Yokoyama, Phys. Rev. D81(2010)104019.
 \bibitem{73} Ya.B. Zel’dovish  and  I.D. Novikov, Sov. Astron. Lett. 10(1967)602.
 \bibitem{74} S. Hawking, MNRAS 152(1971)75
 \bibitem{75} K. Kohri  and  T. Terada Physics Letters B 813 (2021) 136040, arXiv 2009.11853.
 \bibitem{76} V.De Luca, G.Franciolini,  and  A. Riotto Phys. Rev. Lett. 126 (2021)041303, arXiv 2009.08268.
 \bibitem{77} V. Vaskonen  and  H. Veermäe, Phys. Rev. Lett. 126 (2021) 051303, arXiv 2009.07832.
 \bibitem{79} B.J. Carr, Astrophys. J. 201(1975)1.
 \bibitem{78} W.H. Press  and  P. Schechter, Astrophys. J. 187 (1974) 425.
 \bibitem{80} T. Suyama  and  S. Yokoyama, PTEP 2020(2020)023E03.
 \bibitem{81} C. Germani  and  R.K. Sheth, Phys. Rev. D101(2020)063520.
 \bibitem{82} K. Ando, K. Inomata,  and  M. Kawasaki, Phys. Rev. D 97 (2018)103528.
 \bibitem{83} K. Tokeshi, K. Inomata,  and  J. Yokoyama, JCAP 12 (2020)038.
 \bibitem{84} M. Kamionkowski, A. Kosowsky,  and  M.S. Turner, Phys. Rev. D 49(1994)2837.
 \bibitem{85} A. Addazi, Y.F. Cai, Q. Gan, A. Marciano  and  K. Zeng, 2009.10327
 \bibitem{86} Y. Nakai, M. Suzuki, F. Takahashi  and  M. Yamada, Phys. Lett. B 816(2021)136238. 
 \bibitem{87} A.R. Pol et al., Phys. Rev. D 102(2020)083512.
 \bibitem{88} A.Neronov, A.R. Pol, C. Caprini,  and  D. Semikoz, Phys. Rev. D 103 (2021)041302.
 \bibitem{89} R.M. Shannon et al., Science 349(2015)1522.



\end{thebibliography}


\end{document}